\documentclass[12pt]{article}
\usepackage[a4paper,width=170mm,top=15mm,bottom=20mm]{geometry}
\usepackage{newtxmath}

\DeclareMathOperator*{\argminA}{arg\,min}
\usepackage{graphicx}
\usepackage{subcaption}
\usepackage{amsthm}
\newtheorem{corollary}{Corollary}
\usepackage{natbib}
\usepackage{float}
\usepackage{authblk}
\usepackage{orcidlink}
\usepackage{setspace}
\usepackage{algorithm}
\usepackage{algpseudocode}
\usepackage{xcolor}
\usepackage{hyperref}
\hypersetup{
    colorlinks,
    linkcolor={red!50!black},
    citecolor={blue!50!black},
    urlcolor={blue!80!black}
}

\title{Stability Selection via Variable Decorrelation}
\author[1]{Mahdi Nouraie\orcidlink{0000-0002-4792-4994}}
\author[1]{Connor Smith\orcidlink{0000-0002-3955-4348}}
\author[1,2]{Samuel Muller\orcidlink{0000-0002-3087-8127}\thanks{Address for correspondence: samuel.muller@mq.edu.au}}
\affil[1]{School of Mathematical and Physical Sciences, Macquarie University}
\affil[2]{School of Mathematics and Statistics, The University of Sydney}

\date{}
\begin{document}
\begin{spacing}{1}
\maketitle
\begin{abstract}
\noindent{The Lasso is a prominent algorithm for variable selection. However, its instability in the presence of correlated variables in the high-dimensional setting is well-documented. Although previous research has attempted to address this issue by modifying the Lasso loss function, this paper introduces an approach that simplifies the data processed by Lasso. We propose that decorrelating variables before applying the Lasso improves the stability of variable selection regardless of the direction of correlation among predictors. Furthermore, we highlight that the irrepresentable condition, which ensures consistency for the Lasso, is satisfied after variable decorrelation under two assumptions. In addition, by noting that the instability of the Lasso is not limited to high-dimensional settings, we demonstrate the effectiveness of the proposed approach for low-dimensional data. Finally, we present empirical results that indicate the efficacy of the proposed method across different variable selection techniques, highlighting its potential for broader application. The \texttt{DVS} R package is developed to facilitate the implementation of the methodology proposed in this paper.
}
\end{abstract}
\end{spacing}

\section{Introduction}\label{s1}
\label{sec:intro}
\sloppy
Linear regression is formally defined as $\boldsymbol{Y} =  X\boldsymbol{\beta} + \boldsymbol{\varepsilon}$, where $\boldsymbol{Y}$ denotes the $n-$vector of the response variable, $X \in \operatorname{mat}(n,p)$ denotes the fixed design matrix,  $\boldsymbol{\beta}$ denotes the $p-$vector of regression coefficients, and $\boldsymbol{\varepsilon}$ denotes the $n-$vector of random errors. The intercept term is omitted here because, without loss of generality, we assume that the response vector is centred, that is, $\sum_{i = 1}^{n}\boldsymbol{Y}_{i} = 0$. The least absolute shrinkage and selection operator \citep[Lasso;][]{tibshirani1996regression} is a prominent variable selection method that achieves sparsity by adding an $\ell_1-$norm to the least-squares loss function. The regression coefficients of the Lasso are estimated through
\begin{equation}\label{eq1-lasso}
\boldsymbol{\hat{\beta}}(\lambda) = \argminA_{\boldsymbol{\beta} \in \mathbb{R}^{p}} \left(\|\boldsymbol{Y} - X\boldsymbol{\beta}\|_{2}^{2} + \lambda \|\boldsymbol{\beta}\|_{1}\right),
\end{equation}
where $\lambda \in \mathbb{R}^{+}$ denotes the Lasso regularisation parameter.

Variable selection, following the terminology of \citet{meinshausen2010stability}, refers to categorising variables into two distinct groups: the signal group $S := \{k \neq 0 \mid \beta_k \neq 0\}$ and the noise group $N := \{k \neq 0 \mid \beta_k = 0\}$, where $S \cap N = \emptyset$. The primary objective of variable selection is to accurately identify the signal group $S$. When using the Lasso, the set of non-zero coefficients $\hat{S}(\lambda) := \{k \neq 0 \mid \hat{\beta}_{k}(\lambda) \neq 0\}$, can be derived from solving Equation~\eqref{eq1-lasso} through convex optimisation.

The stability of a variable selection method is related to its ability to consistently identify the same variables across different training sets sampled from the same underlying distribution \citep{kalousis2007stability}. This property serves as an indicator of the reproducibility and generalisability of the selection results and has received significant attention in recent years (see, e.g., \citet{meinshausen2010stability}, \citet{nogueira2018stability}, and \citet{JMLR:v25:23-0536}). In the high-dimensional setting, where $p \gg n$, the presence of correlated variables within $X$ can directly affect the stability of the variable selection carried out by Lasso \citep{10.1093/bioinformatics/bty750}. To be more specific, when two variables are highly correlated and their effect sizes are unconstrained, they exhibit a nearly identical effect on the response variable \citep{10.1093/bioinformatics/bty750}. In such cases, Lasso tends to select at random only one of these variables \citep{xu2011sparse, 10.1093/bioinformatics/bty750}. In general, if an irrelevant variable is highly correlated with the relevant variables, Lasso may not be able to distinguish it from the relevant variables with any amount of data and
any amount of regularisation \citep{a1294e48-ce43-314c-867d-a032d7baf484, zhao2006model}.

In this paper, we propose to decorrelate predictor variables before performing variable selection by Lasso. The proposed variable decorrelation process transforms the predictors into an orthonormal basis aiming to reduce noise from inter-variable dependencies and improve interpretability. We show that this transformation improves the stability of variable selection for the Lasso. Although the orthonormal design matrix has long been a subject of interest (see, e.g., \citet{tibshirani1996regression} and \citet{yuan2006model}), to the best of our knowledge, this is the first paper to propose addressing the instability of the Lasso through an orthonormalisation process.  We further show that the irrepresentable condition \citep{zhao2006model, meinshausen2006high}, which guarantees consistent variable selection by the Lasso, holds after variable decorrelation under two assumptions. In addition, we highlight that the instability of the Lasso is not confined to high-dimensional settings, and variable decorrelation can serve as an effective strategy for achieving stable variable selection in low-dimensional data as well.

Several previous studies have attempted to improve the stability of the Lasso by modifying its loss function to better handle correlated variables (see, e.g., \citet{Chen_Ding_Luo_Xie_2013}, \citet{pmlr-v84-takada18a}, and \citet{10.1093/bioinformatics/bty750}). A complementary approach, stability selection, enhances the stability of variable selection by applying a selection method like the Lasso to multiple random half-samples of the original dataset \citep{meinshausen2010stability, shah2013variable}. Stability selection identifies relevant variables by evaluating their selection frequencies across various randomly drawn sub-samples of observations. In this paper, we use the latter approach to identify stable variables following the decorrelation of the variables.

The Gram–Schmidt process \citep{BJORCK1994297} is a fundamental technique for generating an orthonormal set of vectors that spans the same subspace as the original set. The Gram–Schmidt process is traditionally used to compute the thin QR factorisation \citep{golub2013matrix}, and in recent years, has been applied to variable ranking, screening, and selection problems (see, e.g., \citet{stoppiglia2003ranking}, \citet{Liu12122018}, \citet{gram2}, and \citet{https://doi.org/10.1002/cem.3510}). The Gram–Schmidt process is presented in Algorithm \ref{alg-Gram} in the appendix. The resulting matrix $Q$ in Algorithm \ref{alg-Gram} has the same dimensions as the input matrix $X$, with its columns mutually orthogonal and normalised to have an $\ell_2$-norm of one. In this paper, we show that using $Q$, obtained by applying Algorithm \ref{alg-Gram} to the design matrix $X$, as a surrogate for $X$ in variable selection results in greater stability; however, we first comment on its limitations and then focus on refinements 

Algorithm \ref{alg-Gram} sequentially orthogonalises columns of $X$ from index 1 to $p$, giving earlier variables greater influence over the resulting basis when correlations are non-zero. This creates a competitive dynamic, where later correlated variables must offer additional information to be represented. Here, we present a toy example to illustrate the index sensitivity of Algorithm \ref{alg-Gram}. The dataset contains $n = 100$ samples and $p = 5$ variables, with $X_2 = 0.8 X_1$ introducing correlation. The response $\boldsymbol{Y}$ is generated from $\boldsymbol{\beta} = (1, 0, 0, 0, 0)^\top$ with standard normal noise. We apply the Lasso to the orthonormal matrix $Q$ in the two cases: once obtained from the original design matrix, and once from the reordered design matrix after swapping $X_1$ and $X_2$. In both cases, only the first coefficient is non-zero, highlighting how the order of the variables affects the selection results. The corresponding implementation code is available on GitHub\footnote{\url{https://github.com/MahdiNouraie/Variable-Decorrelation/blob/main/Code/Toy_example.R}}. Section \ref{s4} presents further simulations showing that applying Lasso to $Q$ often yields poor selection accuracy. To address this, we propose reordering columns before applying the Gram–Schmidt process. Therefore, in this paper, we introduce a method which involves three key steps: (i) reordering the variables, (ii) applying Gram-Schmidt orthogonalisation, and (iii) performing variable selection on $Q$ using the stability selection with the Lasso.

The rest of the paper is organised as follows. Section \ref{s2} presents and justifies our method. Section \ref{s3} describes the synthetic and real datasets used in the paper. Section \ref{s4} presents the numerical results, and Section \ref{s5} discusses related work and situates the proposed approach within its broader context and potential impact. The appendix contains the Gram-Schmidt process algorithm, proofs of the corollaries, and supplementary numerical results.
%
%
%
%
\section{Methodology}\label{s2}
In this section, we propose a method that reorders the columns of $X$, decorrelates the variables by projecting them onto an orthonormal basis, and integrates the resulting orthonormal matrix into the stability selection framework to improve the selection stability of the Lasso in the presence of correlated predictors.

We consider a dataset $\mathcal{D} := \{(\boldsymbol{x}^\top_{i}, y_{i})\}_{i=1}^{n}$ where each element of $\mathcal{D}$ consists of a univariate response $y_{i} \in \mathbb{R}$ and a $p$-dimensional fixed variable vector $\boldsymbol{x}^\top_{i} \in \mathbb{R}^{p}$. We assume that the columns of the design matrix $X$ are centred and scaled by their means and standard deviations. Following the basic assumptions of linear regression, it is also assumed that $E(\boldsymbol{Y}\mid X)$ is linear in the coefficients, $E(\boldsymbol{\varepsilon}\mid X) = 0$, $\text{Var}(\boldsymbol{\varepsilon} \mid X) = \sigma^2 < \infty$, and error terms are uncorrelated with predictor variables. In addition, we assume that the columns of $X$ are not linearly dependent on each other and that the components of $\boldsymbol{\varepsilon}$ are independent of each other.

Given that Algorithm \ref{alg-Gram} is sensitive to the order of column indices, we propose that the columns of the design matrix first be sorted according to their predictive power for the response variable. To achieve this, we advocate for the use of sure screening methods, given their favourable asymptotic properties, as an appropriate approach for ordering the variables prior to the decorrelation and selection process. Specifically, we employ Air-Holp \citep{AirHOLP}, a data-adaptive variant of ridge high-dimensional ordinary least-squares projection \citep[Ridge-HOLP;][]{10.1111/rssb.12127}. Although other screening and ranking methods may be used, comparing the efficacy of different ranking methods is beyond the scope of this paper. The variables should be ordered according to the output of the chosen ranking method, ensuring that the variable with the highest predictive power for the response variable is placed in the first column, etc. In the next step, we decorrelate the variables within the ordered design matrix by projecting them onto a new space of orthonormal vectors by Algorithm \ref{alg-Gram}. Then, we propose using $Q$ and $\boldsymbol{Y}$ in the stability selection using the Lasso to select the relevant variables. The general method is outlined in Algorithm \ref{alg-pipeline}.

\begin{algorithm}
\caption{Stability Selection via Variable Decorrelation}
\label{Alg1}
\begin{algorithmic}[1]
    \State \textbf{Input:} Design matrix \( X \), response vector \( \boldsymbol{Y} \)
    \State \textbf{Output:} Relevant variables 

    \State \textbf{Step 1:} Sort the columns of \( X \) using Air-HOLP to prioritise variables with higher predictive power for the response variable.
    
    \State \textbf{Step 2:} Project the ordered design matrix \( X \) onto a new space \( Q \) with orthonormal vectors using Algorithm \ref{alg-Gram} to decorrelate the variables.
    
    \State \textbf{Step 3:} Integrate the transformed design matrix \( Q \) and the response vector \( \boldsymbol{Y} \) into the stability selection framework using the Lasso.
\end{algorithmic}\label{alg-pipeline}
\end{algorithm}

It should be noted that by applying the Lasso to the orthonormal matrix $Q$, the resulting solution has a closed-form expression given by $\hat{\beta}_j = \text{sign}(\hat{\beta}_j^{0}) ( |\hat{\beta}_j^{0}| - \lambda )^{+}$, where $\hat{\beta}_j^{0}$ denotes the full least-squares estimate \citep{tibshirani1996regression}.

\subsection*{Justifying the Surrogate}
\subsubsection*{Why columns of $Q$ are good proxies for columns of $X$?}
As shown in Algorithm \ref{alg-Gram}, the $j$th column of $R$ contains the normalising factor for the corresponding orthogonal vector in $Q$ on the diagonal, with zeros below it. The entries above the diagonal capture the linear dependencies between $X_j$ and the previously orthonormalised vectors. Therefore, we conclude that the component of $X_j$ contributing uniquely to the span of $X$ is retained in its orthonormalised form as $Q_j$. This enables us to use $Q$ as a suitable surrogate for variable selection compared to $X$ when the variables are correlated.

\subsubsection*{Is the principle of resampling from observations preserved when using $Q$ in stability selection?}

Stability selection relies on repeatedly drawing random sub-samples of the observations to enhance the stability of variable selection. An important consideration in the context of stability selection is whether it is valid to use the orthonormal matrix $Q$ within the stability selection, given that $Q$ is computed using all observations from the original design matrix $X$?  

For any sub-sample of observations, let $X_{b\bullet}$  represent the corresponding rows of $X$. Since $X = QR$, we have $X_{b\bullet} = Q_{b\bullet} R$. This relationship implies that sub-sampling the rows of $Q$ is intrinsically linked to sub-sampling the original observations in $X$, followed by a fixed transformation determined by $R$. Consequently, the fundamental principle of stability selection over sub-samples of the original data is preserved.

\subsection*{Regularisation Tuning}
In our implementations, we follow the guidelines recently introduced by \citet{nouraie2024selection}, who proposed a method for tuning the Lasso regularisation parameter that balances selection stability and prediction accuracy, consistent with the approach proposed by \citet{10.5555/2567709.2567772}. Specifically, \citet{nouraie2024selection} recommended recording the selection results for each regularisation value $\lambda$ in the corresponding binary selection matrix $M(\lambda) \in \operatorname{mat}(B, p)$, where $B$ denotes the number of sub-samples used when applying the Lasso in the stability selection process. They leveraged the stability measure introduced by \citet{nogueira2018stability} to assess the stability of $M(\lambda)$ and determine the optimal regularisation value. 
The stability measure is defined as

\begin{equation}\label{eqn: phi}
    \hat{\Phi}(M(\lambda)) := 1 - \frac{\frac{1}{p}\sum_{j=1}^{p} s_{j}^{2}}{\frac{q(\lambda)}{p}\left(1 - \frac{q(\lambda)}{p}\right)};\quad \lambda \in \Lambda,
\end{equation}
where $s_{j}^{2}$ denotes the unbiased sample variance of the binary selection statuses of the $j$th variable within $M(\lambda)$, $q(\lambda)$ denotes the average number of variables selected under the regularisation parameter $\lambda$, and $\Lambda$ denotes the set of candidate regularisation values. \citet{nouraie2024selection} proposed $\lambda_{\text{stable}}$, defined as the smallest regularisation value for which the corresponding stability exceeds 0.75, and $\lambda_{\text{stable-1sd}}$, defined as the smallest regularisation value for which the corresponding stability remains within one standard deviation of the maximum stability achieved. $\lambda_{\text{stable}}$ is identified as an optimal choice on the Pareto front \citep{pareto1896cours} of selection stability and prediction accuracy, while $\lambda_{\text{stable-1sd}}$ is recommended when $\lambda_{\text{stable}}$ is not attainable.

\subsection*{Consistent Variable Selection}
\citet{zhao2006model} and \citet{meinshausen2006high} introduced the irrepresentable condition for the Lasso, which ensures the consistent variable selection. This condition can be mathematically formulated as $\|X_{N}^\top X_{S} (X_{S}^\top X_{S})^{-1}\|_{\infty} < 1$, where $X_{N}:= [X_{j} \mid j \in N]$ represents the sub-matrix of $X$ that contains only noise variables, and $X_{S}:= [X_{j} \mid j \in S]$ is the sub-matrix that contains only signal variables. Here, we demonstrate that by orthonormalising the design matrix, this condition is satisfied under two assumptions.

\begin{corollary} If the columns of the design matrix $X$ are not linearly dependent on each other, and if the variable ordering method positions all relevant variables in earlier columns relative to the irrelevant variables that are correlated with them, the irrepresentable condition for the Lasso is satisfied by employing the matrix $Q$, which represents the orthonormal matrix obtained through applying Algorithm \ref{alg-Gram} to the ordered $X$, as the new design matrix. \label{Corollary-consistency}\end{corollary}

The proof of Corollary \ref{Corollary-consistency} is shown in the appendix. The first assumption in Corollary \ref{Corollary-consistency} ensures that division by zero is avoided during the normalisation step in Algorithm \ref{alg-Gram}. The second assumption pertains to the effectiveness of the ranking method employed. For Ridge-HOLP, it is established in \citet{10.1111/rssb.12127} that the method is screening consistent—that is, if a model of the same size as the true model is retained, it asymptotically coincides with the true model. However, the requirement in Corollary \ref{Corollary-consistency} is weaker. We do not require that all relevant variables precede all irrelevant ones in the ranking; rather, it suffices that each relevant variable precedes those irrelevant variables that are correlated with it. According to \citet{10.1111/rssb.12127}, a critical condition for the screening consistency of Ridge-HOLP is that the condition number of the covariance matrix of the design matrix satisfies $\kappa(\Sigma) \leq c n^{\tau}$, where $c > 0$ and $0 \leq \tau < 1/7.5$. Here, $\kappa(\Sigma) = \lambda_{\max}(\Sigma) / \lambda_{\min}(\Sigma)$ denotes the condition number of $\Sigma$, and $\lambda_{\max}(\Sigma)$ and $\lambda_{\min}(\Sigma)$ denote the largest and smallest eigenvalues of $\Sigma$ respectively. One instance in which this condition is met is when the groups of correlated variables are mutually independent, each group contains finitely many variables, and the variables within each group exhibit a compound symmetric structure with a correlation value less than one. The assumptions required for the screening consistency of Ridge-HOLP are stated in Theorem 3 of \citet{10.1111/rssb.12127}.

\subsection*{Stable Variable Selection}
Although the instability of Lasso in the presence of correlated variables has been recognised for years, \citet{faletto2022cluster} formally demonstrated that
in the context of a simple linear model with two highly correlated, equally relevant variables and one less important variable, as $n$ approaches infinity, the selection frequencies of the two more relevant variables tend to 0.5, while the selection frequency of the less relevant variable tends to 1 when stability selection is applied using Lasso. Building on this, we now demonstrate how our method improves selection stability.

\begin{corollary}
Assume that under the conditions specified in Theorem 1 and Corollary 2 of \citet{faletto2022cluster}, within the design matrix $X \in \operatorname{mat}(n,3)$, there exist two highly correlated and equally relevant variables, denoted $X_1$ and $X_2$, which are uncorrelated with $X_3$. Let the corresponding regression coefficients be $\beta_1 = \beta_2 > \beta_3 = 1$. When stability selection with Lasso is applied to the orthonormal matrix $Q$, obtained by applying Algorithm~\ref{alg-Gram} to $X$, it results in higher selection stability compared to applying stability selection with Lasso directly to $X$. This improvement in stability is evident when the stability measure defined in Equation~\eqref{eqn: phi} is used to evaluate stability. \label{Corollary-stability}
\end{corollary}
The proof of Corollary \ref{Corollary-stability} is shown in the appendix. As is evident, the ordering step is not involved in Corollary \ref{Corollary-stability}. As shown in Corollary \ref{Corollary-consistency}, this step affects selection consistency rather than stability. In addition, since Corollary \ref{Corollary-stability} is established in a low-dimensional setting, we conclude that instability, although often discussed in high-dimensional contexts, is also a concern in low-dimensional settings due to the role of correlation. The numerical results supporting this conclusion are provided in Section \ref{s4}. It should also be noted that, as argued by \citet{nogueira2018stability}, any proper stability measure should be a strictly decreasing function of the variance of the selection status of the variables. Therefore, Corollary \ref{Corollary-stability} holds for any proper stability measure, not only for the specific measure defined in Equation~\eqref{eqn: phi}.

\section{Datasets}\label{s3}
We evaluated our method through experiments on both synthetic and real bioinformatics datasets, as presented below.

\subsection*{Synthetic Data}

As synthetic data, we consider a dataset with a sample size of $n = 50$ that includes $p = 500$ predictor variables. For each dataset, the predictor variables $\boldsymbol{x}^\intercal_{i}$ are independently drawn from the Normal distribution $\mathcal{N}(\boldsymbol{0}, \Sigma)$, where 
$
\Sigma \in \operatorname{mat}(p,p)$ has a diagonal of ones, and the off-diagonal elements are defined as follows
$$
\sigma_{jk} =
\begin{cases}  
\rho_g, & \text{if } j, k \text{ belong to the same group } G_g, \text{ where } g = 1, \dots 5,  \\  
0, & \text{otherwise;}  
\end{cases}  
$$
where each group consists of consecutive indices
$$
G_g = \left\{ \frac{(g-1)p}{5} + 1, \dots \frac{gp}{5} \right\}, \quad g = 1, \dots 5.
$$
Each group $G_g$ is associated with a distinct covariance parameter $\rho_g$. Specifically, the covariance values are set as $\rho_1 = 0.2$, $\rho_2 = 0.4$, $\rho_3 = 0.6$, $\rho_4 = 0.8$, and $\rho_5 = 0.9$. Within each group, a single variable is designated as active, and we consider two extreme cases: (i) the lowest-indexed variable in each group is active, and (ii) the highest-indexed variable in each group is active. 

The true regression coefficients of active variables are defined as $
\boldsymbol{\beta} = (4, 3.5, 3, 2.5, 2)^{\top}
$ following the order of the groups and there are $99$ inactive variables in each group. The error term $\boldsymbol{\varepsilon}$ is an i.i.d. sample from the standard Normal distribution $\mathcal{N}(0,1)$. 

In Section \ref{s4}, we extend the initial data generation setting to explore two additional scenarios aimed at deepening our understanding of the behaviour of the selection method. First, we investigate the impact of altering the sign of the covariance values to assess how the procedure responds to changes in the direction of the variable associations. Second, we examine a reduced-dimensional setting in which the number of variables is decreased from 500 to 20, while maintaining the original group structure, to evaluate the performance of our method in a low-dimensional context.

\subsection*{Riboflavin Data}

For the first real example, we use the well-established `Riboflavin' dataset, which focusses on the production of riboflavin (vitamin B2) from various Bacillus Subtilis. This dataset, provided by the Dutch State Mines Nutritional Products, is accessible via \texttt{hdi} R package \citep{hdi-package}. It comprises a single continuous response variable that represents the logarithm of the riboflavin production rate, alongside $p = 4,088$ variables, which correspond to the logarithm of expression levels for 4,088 bacterial genes. The primary objective of analysing this dataset is to identify genes that are associated with riboflavin production, with the ultimate goal of genetically engineering bacteria to enhance riboflavin yield. Data were collected from $n = 71$ relatively homogeneous samples, which were repeatedly hybridised during a feed-batch fermentation process involving different engineered strains and varying fermentation conditions. \citet{buhlmann2014high} employed stability selection with Lasso and identified three genes—\texttt{LYSC\_at}, \texttt{YOAB\_at}, and \texttt{YXLD\_at}—as relevant genes. Later \citet{nouraie2024selection} examined this dataset using stability selection and Lasso and showed that the stability of results is very low since $\lambda_{\text{stable}}$ does not exist and $\hat{\Phi}(M(\lambda_{\text{stable-1sd}}))$ converges to around $0.2$. 

\subsection*{Affymetrix Rat Genome 230 2.0 Array}

As an additional real-world example, we investigate `Affymetrix Rat Genome 230 2.0 Array' microarray data introduced by \citet{scheetz2006regulation}. This dataset comprises $n = 120$ twelve-week-old male rats, with expression levels recorded for nearly 32,000 gene probes for each rat. The primary objective of this analysis is to identify the probes most strongly associated with the expression level of the TRIM32 probe (\texttt{1389163\_at}), which has been linked to the development of Bardet-Biedl syndrome \citep{chiang2006homozygosity}. This genetically heterogeneous disorder affects multiple organ systems, including the retina. In accordance with the pre-processing steps outlined by \citet{huang2008adaptive}, we excluded gene probes with a maximum expression level below the $25$th percentile and those exhibiting an expression range smaller than $2$. This filtering process yielded a refined set of $p = 3,083$ gene probes that demonstrated sufficient expression and variability for further analysis. \citet{nouraie2024selection} analysed this dataset using stability selection and Lasso and showed that the stability of the results is very low since $\lambda_\text{stable}$ does not exist and $\hat{\Phi}(M(\lambda_{\text{stable-1sd}}))$ converges to around $0.15$.

\section{Numerical Results}\label{s4}

In this section, we illustrate the efficacy of our method through its application to synthetic and real data introduced in Section \ref{s3}. 
We used the corresponding GitHub repositories of \citet{AirHOLP} and \citet{nogueira2018stability} to execute Air-HOLP and apply the stability measure of Equation~\eqref{eqn: phi} respectively. For regularisation tuning, we used the GitHub repository associated with \citet{nouraie2024selection}. To create a set of candidate regularisation values $\Lambda$, we use a $10$-fold cross-validation using the \texttt{cv.glmnet} function from the \texttt{glmnet} R package \citep{friedman2010regularization} on the entire dataset $\mathcal{D}$. The Air-HOLP method takes an input parameter, \texttt{Threshold}, which specifies the number of variables to retain during the screening process. In all experiments, this parameter was fixed at 10 to eliminate its influence on the analysis. In addition, the maximum number of iterations for Air-HOLP was set to 10 across all experiments. To implement the Gram-Schmidt process, we used the \texttt{grahm\_schimdtR} function, which was sourced from a solution shared on the Stack Overflow platform\footnote{\url{https://stackoverflow.com/questions/15584221/gram-schmidt-with-r}}.

\subsection*{Synthetic Data}
We generate 100 datasets with varying random seeds, following the setting described in Section \ref{s3}. For each dataset we set the number of sub-samples $B = 100$. We keep this setting of number of datasets and number of sub-samples over all synthetic experiments. In our implementation, we use $\lambda_{\text{stable-1sd}}$ for each dataset, as $\lambda_{\text{stable}}$ may not exist for some datasets. This approach ensures consistent regularisation tuning across all datasets.

\begin{figure}[H]
    \centering
    \subfloat[First scenario \label{fig1:fig1}]{
        \includegraphics[width=0.45\textwidth]{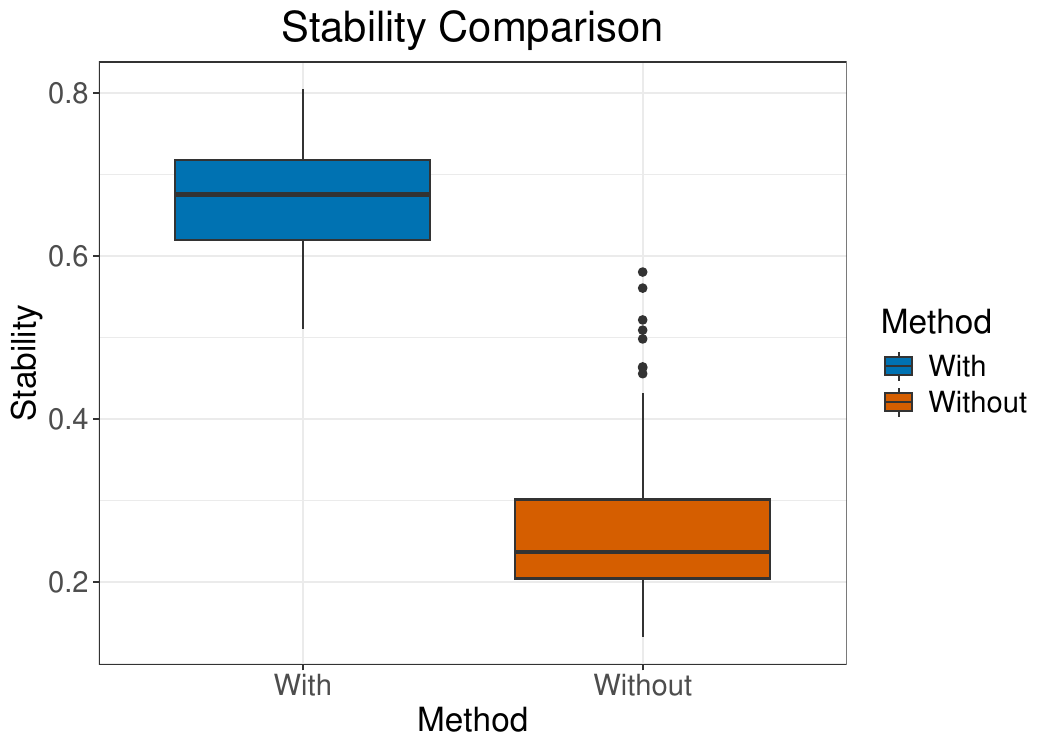}
    }
    \hfill
    \subfloat[First scenario \label{fig1:fig2}]{
        \includegraphics[width=0.45\textwidth]{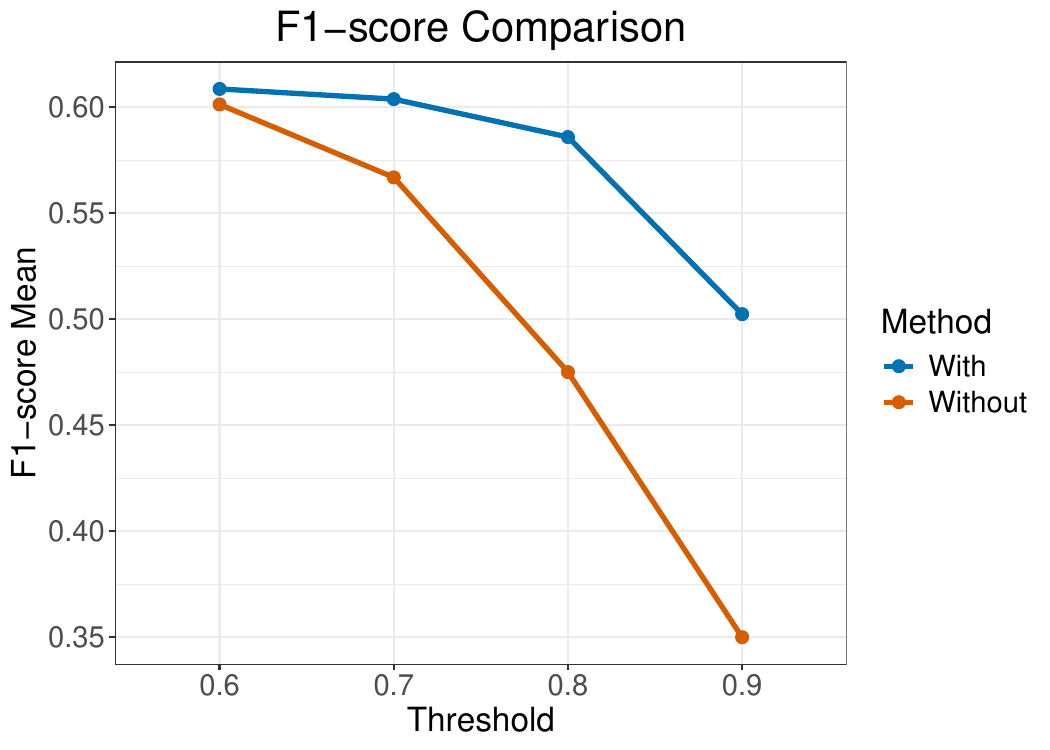}
    }
    \\
    \subfloat[Second scenario \label{fig1:fig3}]{
        \includegraphics[width=0.45\textwidth]{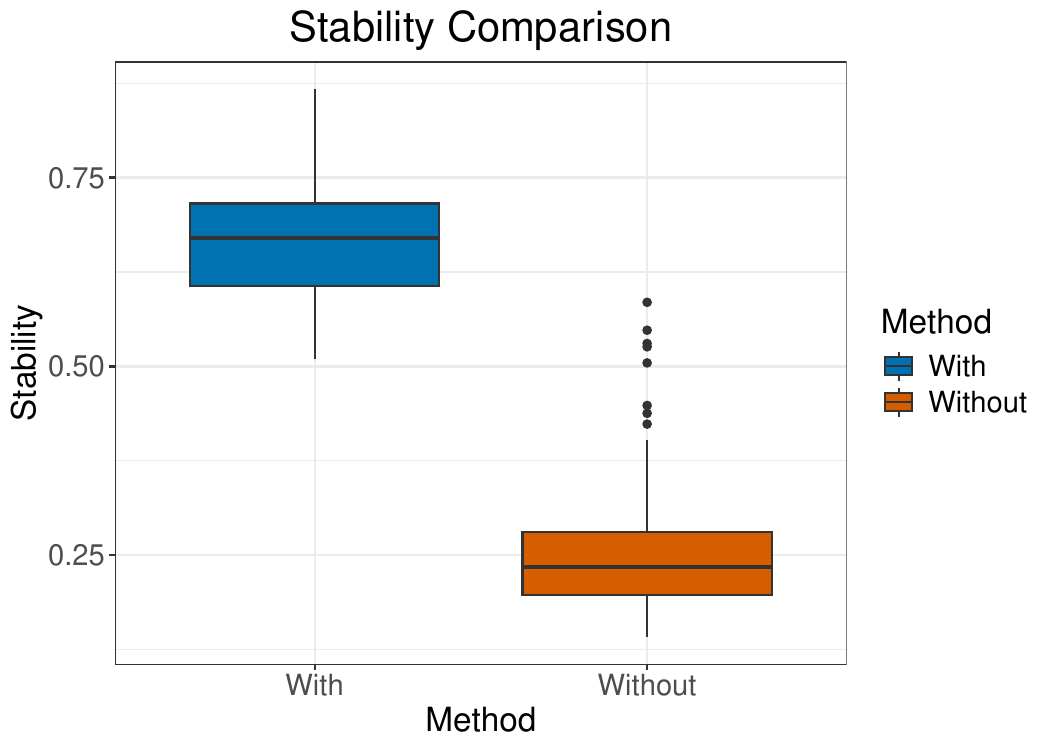}
    }
    \hfill
    \subfloat[Second scenario \label{fig1:fig4}]{
        \includegraphics[width=0.45\textwidth]{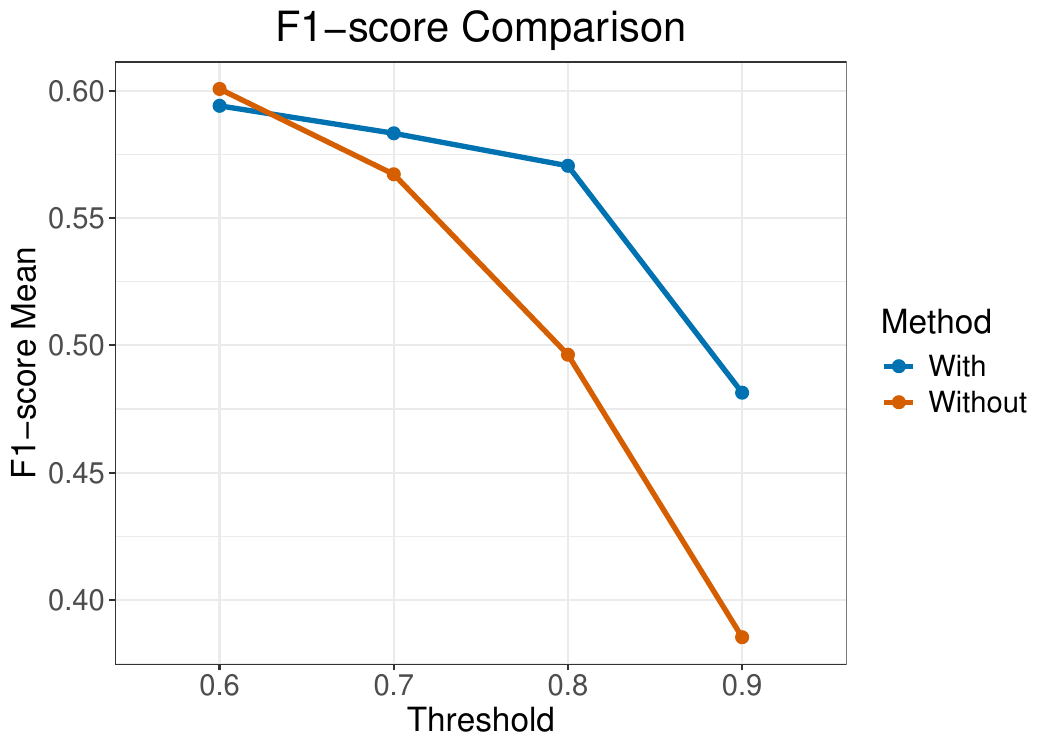}
    }
    \caption{Comparing Stability and F1-score between stability selection using the Lasso with and without the variable decorrelation for simulation scenarios.}
    \label{fig1}
\end{figure}

Figure \ref{fig1} presents a comparative analysis of stability selection using the Lasso, evaluating both selection stability and F1-score, which is the harmonic mean of precision and recall, with and without the implementation of our method. Here, the F1-score is computed to assess selection accuracy, treating the selection task as a binary classification problem. Figures \ref{fig1:fig1} and \ref{fig1:fig2} correspond to the first simulation scenario, whereas Figures \ref{fig1:fig3} and \ref{fig1:fig4} pertain to the second simulation scenario.
As illustrated in Figure \ref{fig1:fig1}, the application of the pre-processing steps significantly enhances stability. Figure \ref{fig1:fig2} depicts the average F1-score of variable selection across different values of the stability selection decision making threshold $\pi_{\text{thr}}$. Figure \ref{fig1:fig2} reveals that, although the primary focus of this paper is on improving stability, our method also yield a notable enhancement in the F1-score. This observation aligns with expectations in light of the asymptotic consistency established in Corollary \ref{Corollary-consistency}.

Figure \ref{fig1:fig3} demonstrates a significant increase in stability, while Figure \ref{fig1:fig4} illustrates a higher average F1-score when employing our method across most of the decision threshold values. This outcome is consistent with the results observed in Figure \ref{fig1:fig1} and \ref{fig1:fig2}. Figures \ref{fig1:fig1} and \ref{fig1:fig3} illustrate that a significant enhancement in stability is observed when correlation is removed from the data. In fact, it is difficult to avoid the conclusion that correlation among predictor variables constitutes a primary source of instability in variable selection.

\subsection*{Why the Ordering Step is Important?}
\begin{figure}[H]
    \centering
    \subfloat[Stability\label{fig2:fig1}]{
        \includegraphics[width=0.45\textwidth, height = 0.4\textwidth]{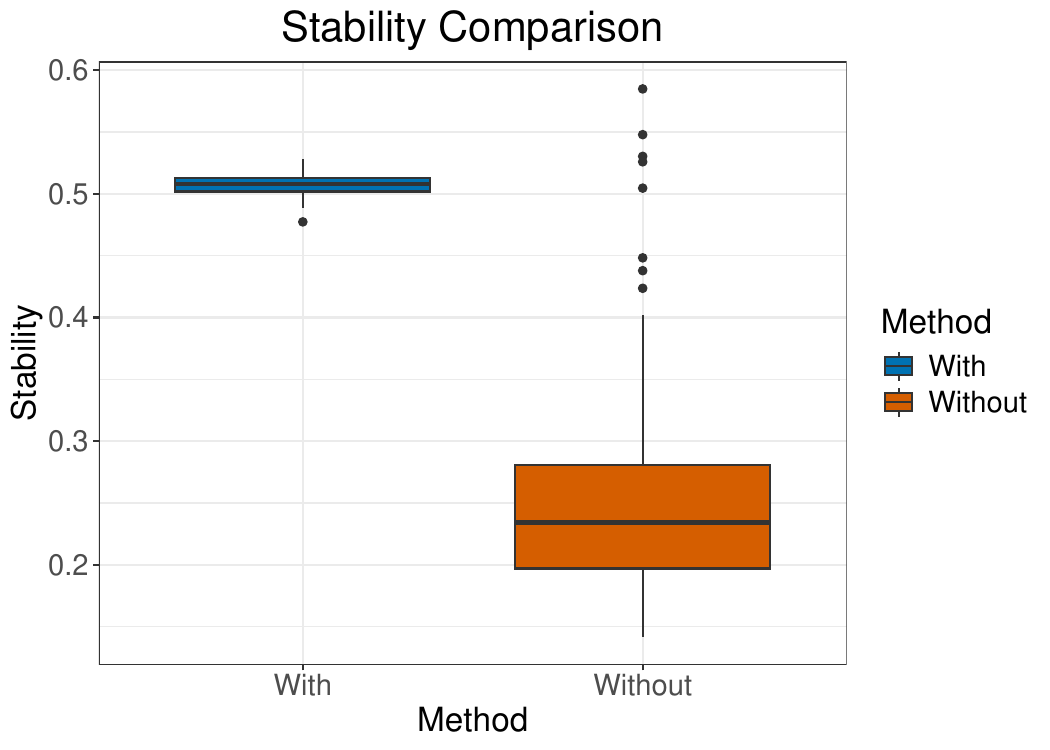}
    }
    \hfill
    \subfloat[F1-score\label{fig2:fig2}]{
        \includegraphics[width=0.45\textwidth, height = 0.4\textwidth]{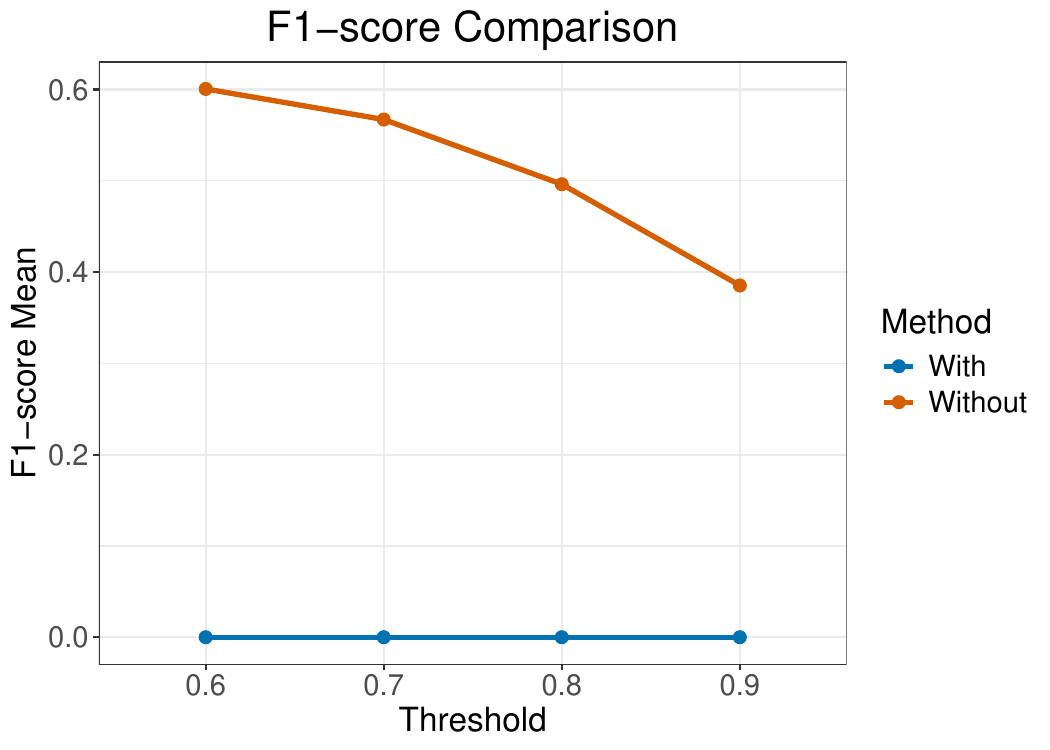}
    }
    \caption{Comparing Stability and F1-score between stability selection using the Lasso with and without the Gram-Schmidt orthonormalisation for the second simulation scenario}
    \label{fig2}
\end{figure}

Based on Corollary \ref{Corollary-consistency}, the ordering step is essential for accurate (consistent) selection. To highlight its importance, we replicate the process for the second simulation scenario, without ordering the variables first. Therefore, we only apply Gram-Schmidt orthonormalisation to the design matrix $X$. Figure \ref{fig2:fig1} shows that this improves selection stability, in line with Corollary \ref{Corollary-stability}; but Figure \ref{fig2:fig2} reveals that the F1-score remains zero at all thresholds. This demonstrates that without ordering, the results may be stable but inaccurate. Therefore, it can be concluded that the ordering step enhances selection accuracy, while the orthonormalisation step contributes to selection stability.

\subsection*{Varying Covariance Signs}
\begin{figure}[H]
    \centering
    \subfloat[negative signs \label{fig3:fig1}]{
        \includegraphics[width=0.45\textwidth]{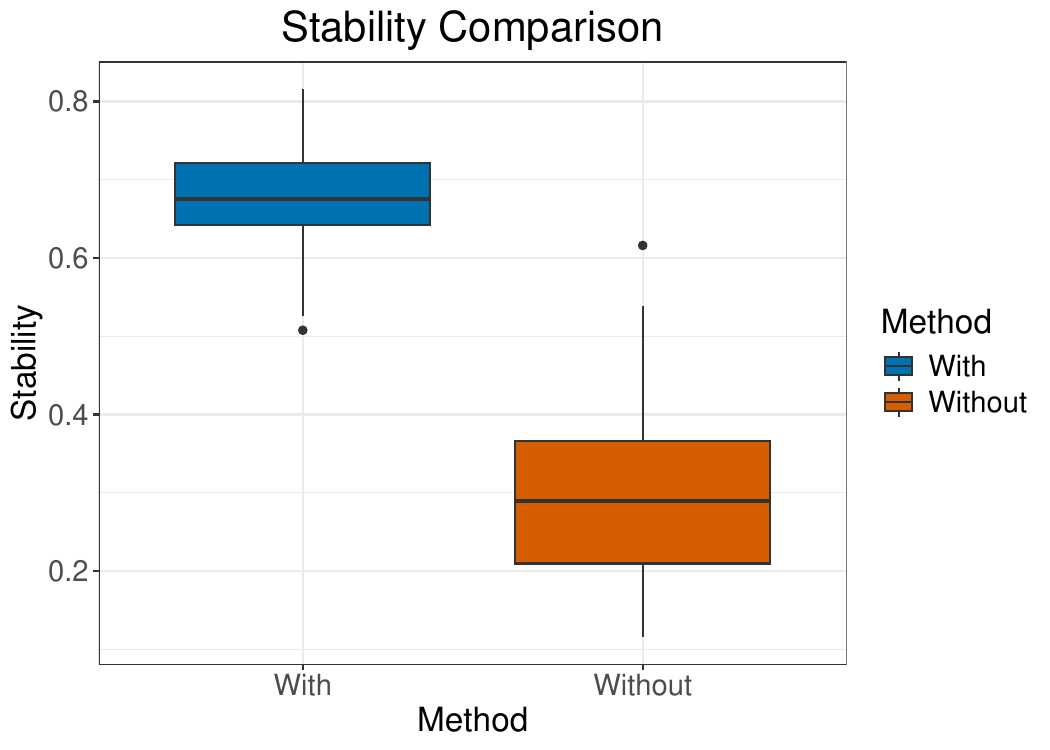}
    }
    \hfill
    \subfloat[negative signs  \label{fig3:fig2}]{
        \includegraphics[width=0.45\textwidth]{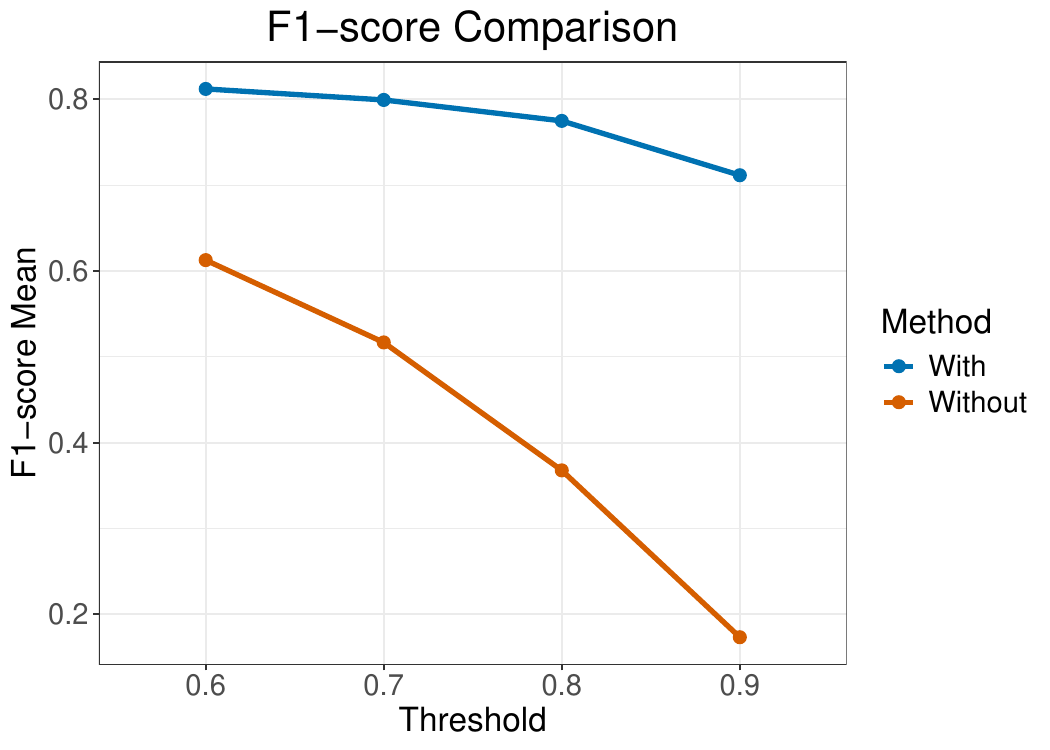}
    }
    \\
    \subfloat[Mixed signs \label{fig3:fig3}]{
        \includegraphics[width=0.45\textwidth]{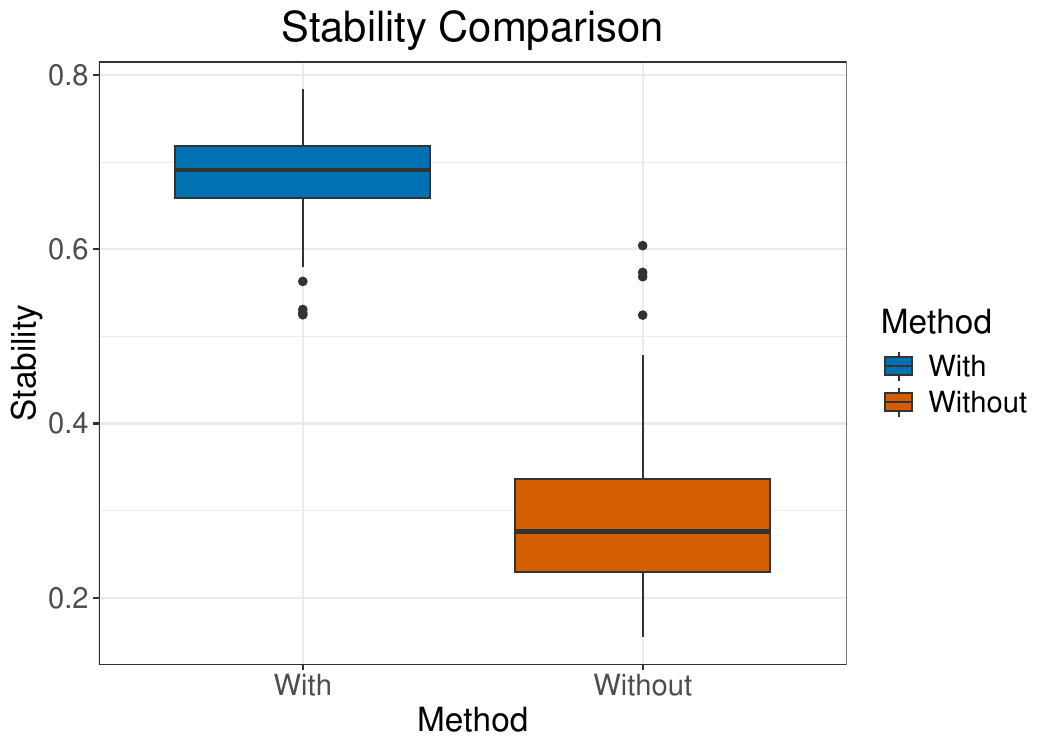}
    }
    \hfill
    \subfloat[Mixed signs \label{fig3:fig4}]{
        \includegraphics[width=0.45\textwidth]{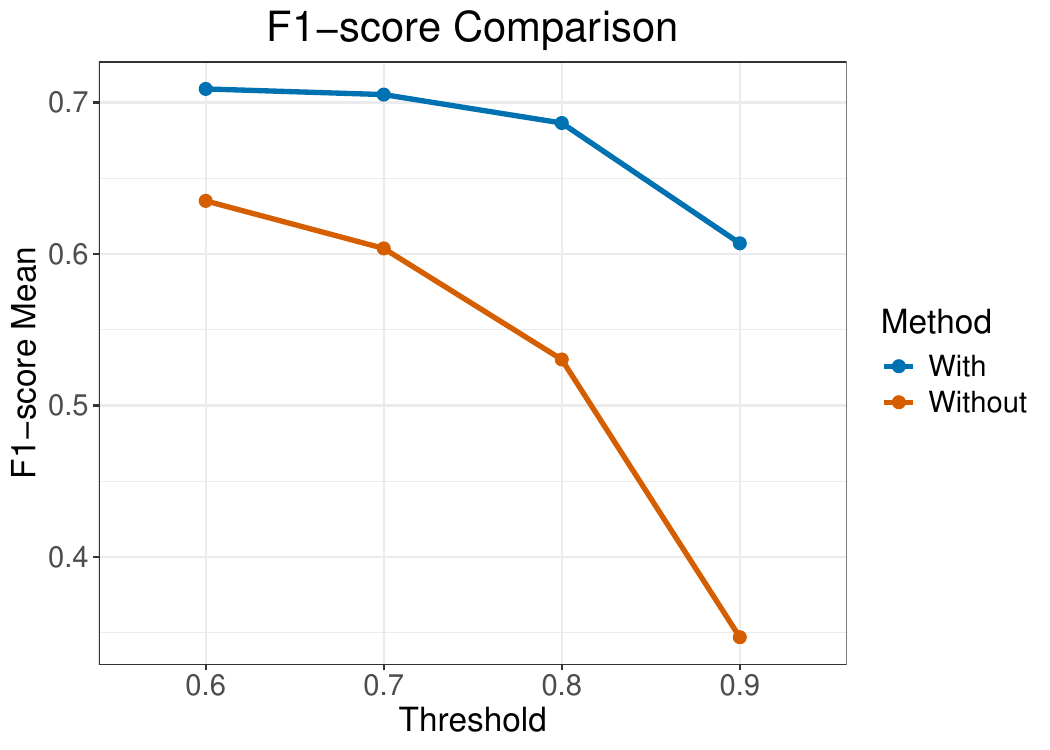}
    }
    \caption{Comparing Stability and F1-score between stability selection using the Lasso with and without the variable decorrelation for the second simulation scenario with varying covariance signs}
    \label{fig3}
\end{figure}

To evaluate the effectiveness of our method in relation to the sign of correlations among variables, we consider the second simulation scenario. However, we alter the sign of the covariance values within correlated groups, specifically setting $\rho_{1} = -0.2$, $\rho_{2} = -0.4$, $\rho_{3} = -0.6$, $\rho_{4} = -0.8$, and $\rho_{5} = -0.9$. We use the \texttt{make.positive.definite} function from the \texttt{corpcor} R package \citep{corpcor} to obtain the nearest positive definite matrix, which is then used as the covariance matrix. Figure \ref{fig3:fig1} and \ref{fig3:fig2} demonstrate that, although the signs of the covariances are reversed, the improvements in both stability and accuracy are maintained.

Figure \ref{fig3:fig3} and \ref{fig3:fig4} show the results regarding a more realistic setting where there is a mix of positive and negative covariance values; specifically, $\rho_{1} =0.2$, $\rho_{2} = -0.4$, $\rho_{3} = 0.6$, $\rho_{4} = -0.8$, $\rho_{5} = 0.9$. Once again, the improvements are retained. In summary, the results from the simulation scenarios demonstrate that our method  enhances both selection stability and accuracy when applying the Lasso to correlated high-dimensional data, regardless of the sign of covariance values. This result is consistent with expectations, as no constraints were imposed on the signs of the covariance values in Section \ref{s2}.

\subsection*{Low-Dimensional Data}
As discussed in Section \ref{s2}, Corollary \ref{Corollary-stability} demonstrates that our method improves selection stability even in low-dimensional settings. To further assess this claim, we modify the data generation process in the second simulation scenario by reducing the number of variables from 500 to 20, while keeping all other conditions unchanged. This adjustment places us in a low-dimensional context comprising five groups of variables, each consisting of four variables. Within each group, the variable with the highest index is designated as the relevant one.

\begin{figure}[H]
    \centering
    \subfloat[Stability\label{fig4:fig1}]{
        \includegraphics[width=0.45\textwidth, height = 0.4\textwidth]{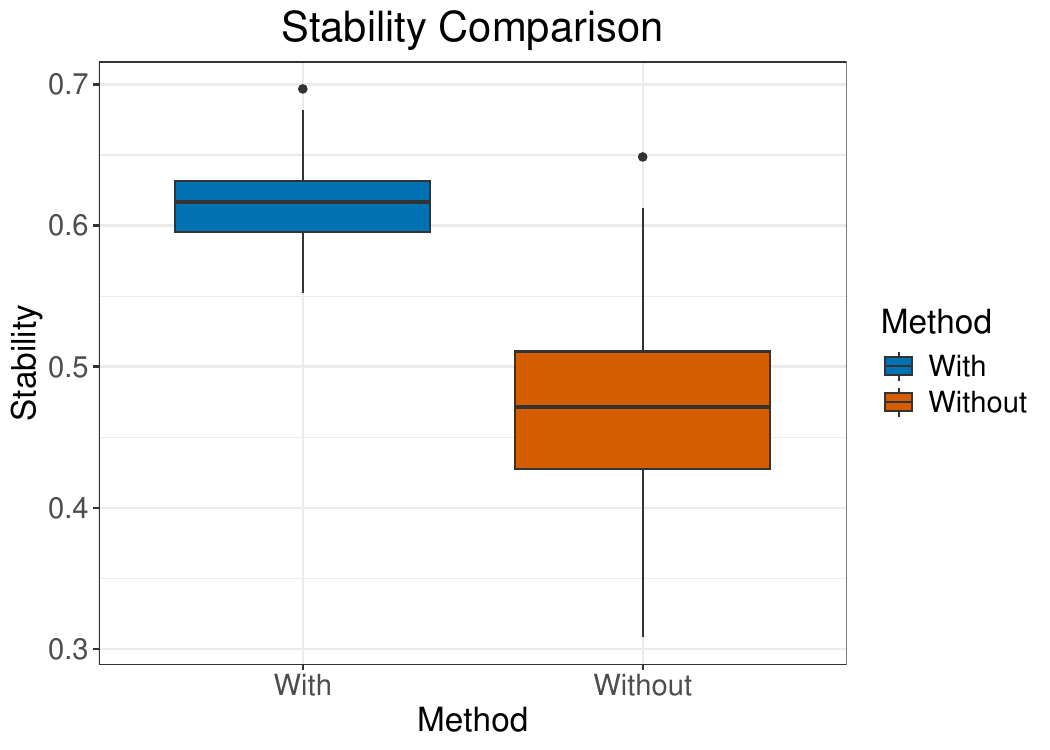}
    }
    \hfill
    \subfloat[F1-score\label{fig4:fig2}]{
        \includegraphics[width=0.45\textwidth, height = 0.4\textwidth]{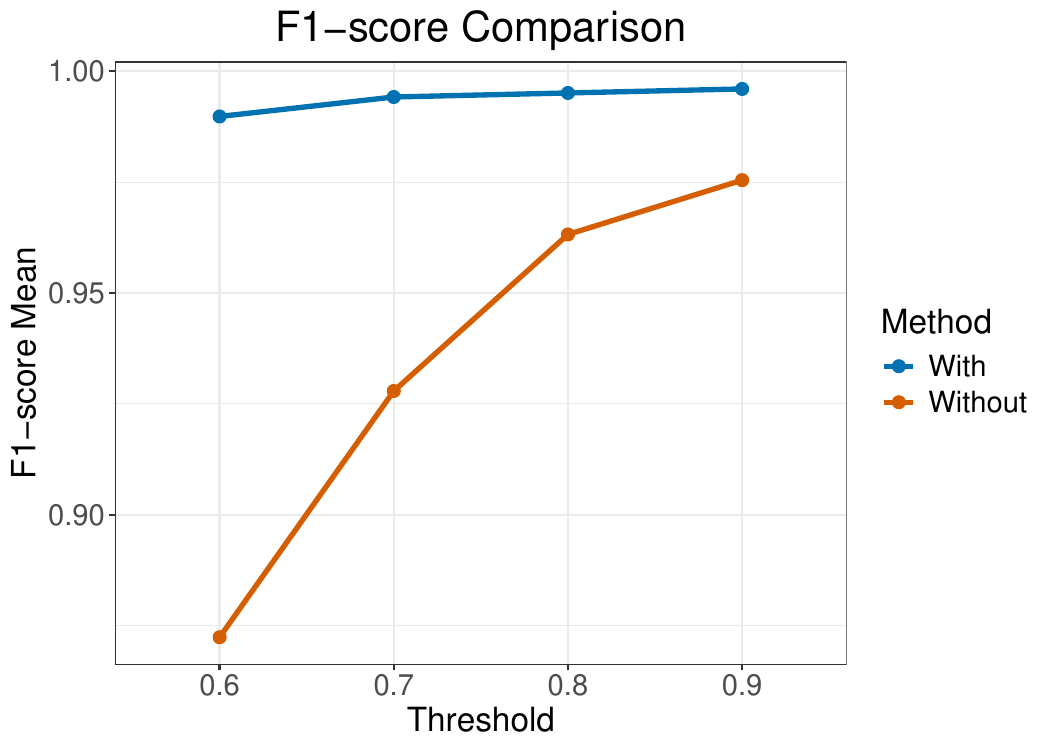}
    }
    \caption{Comparing Stability and F1-score between stability selection using the Lasso with and without the variable decorrelation for the low-dimensional data}
    \label{fig4}
\end{figure}

Figure \ref{fig4:fig1} illustrates that, as expected, our method leads to an improvement in the selection stability. In addition, Figure \ref{fig4:fig2} shows that the variable decorrelation also leads to an improvement in the selection accuracy. The increasing trend in selection accuracy with respect to $\pi_{\text{thr}}$ is in contrast to the patterns observed in the previous figures, highlighting a distinctive behaviour in the low-dimensional setting. In addition, across all figures presented so far on the selection accuracy, it is evident that the selection accuracy obtained using variable decorrelation exhibits greater robustness to changes in $\pi_{\text{thr}}$.

\subsection*{Comparison with Alternative Techniques}
In this section, we compare our method with several regularisation-based variable selection techniques, including Elastic Net \citep[ENet;][]{zou2005regularization}, Smoothly Clipped Absolute Deviation \citep[SCAD;][]{Fan01122001}, Minimax Concave Penalty \citep[MCP;][]{MCP}, and Sparse Group Lasso \citep[SGL;][]{simon2013sparse}. The comparison is conducted under the second simulation scenario. The regularisation parameters were selected using the $\lambda_{\text{stable-1sd}}$ strategy. For the ENet, the mixing parameter \texttt{alpha} was set to 0.2. As the SGL requires pre-defined group indices for the variables, the grouping information was provided accordingly.

\begin{figure}[H]
    \centering \includegraphics[width=0.75\linewidth]{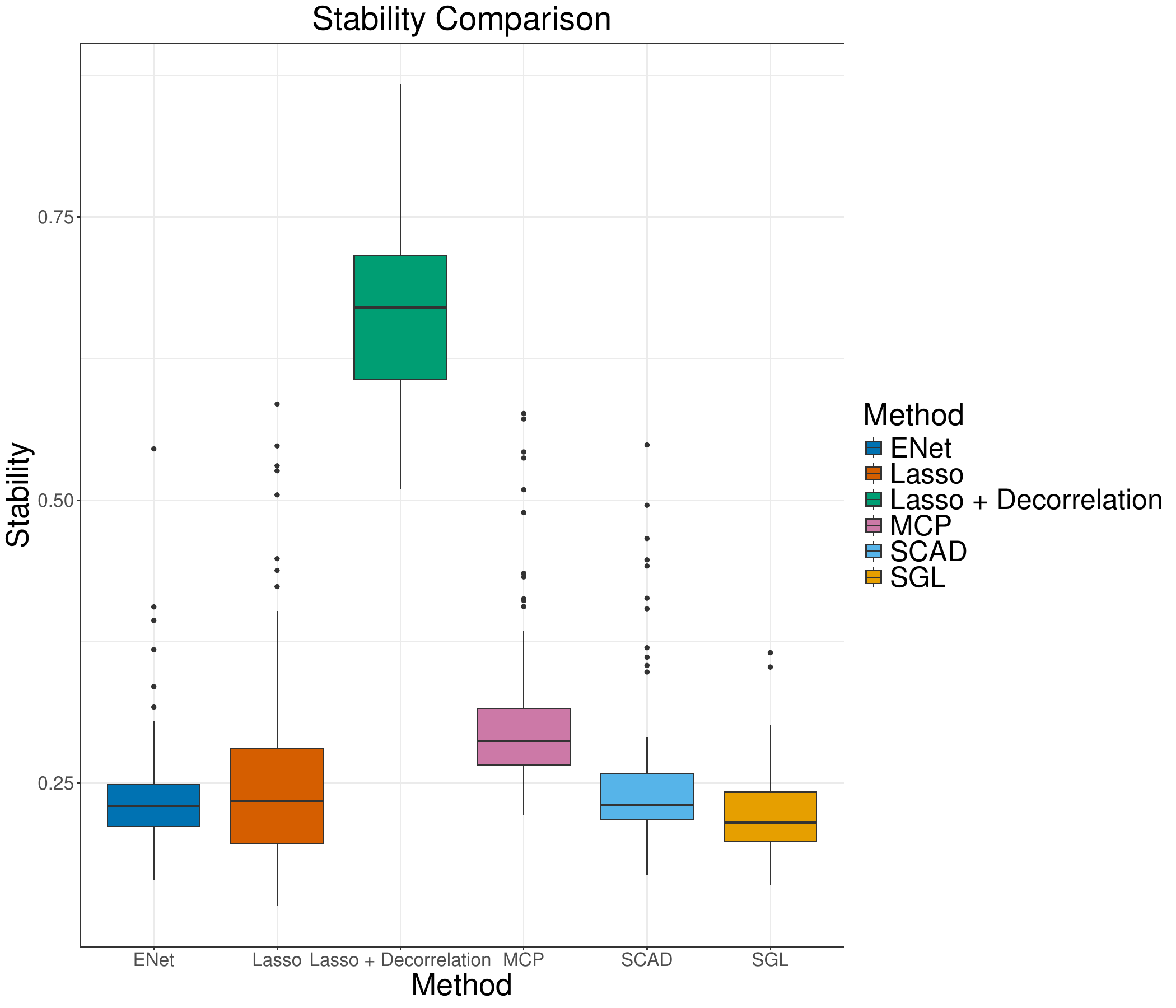}
    \caption{Comparison of selection stability across stability selection with different selection algorithms}
    \label{fig5}
\end{figure}

\begin{figure}[H]
    \centering \includegraphics[width=0.75\linewidth]{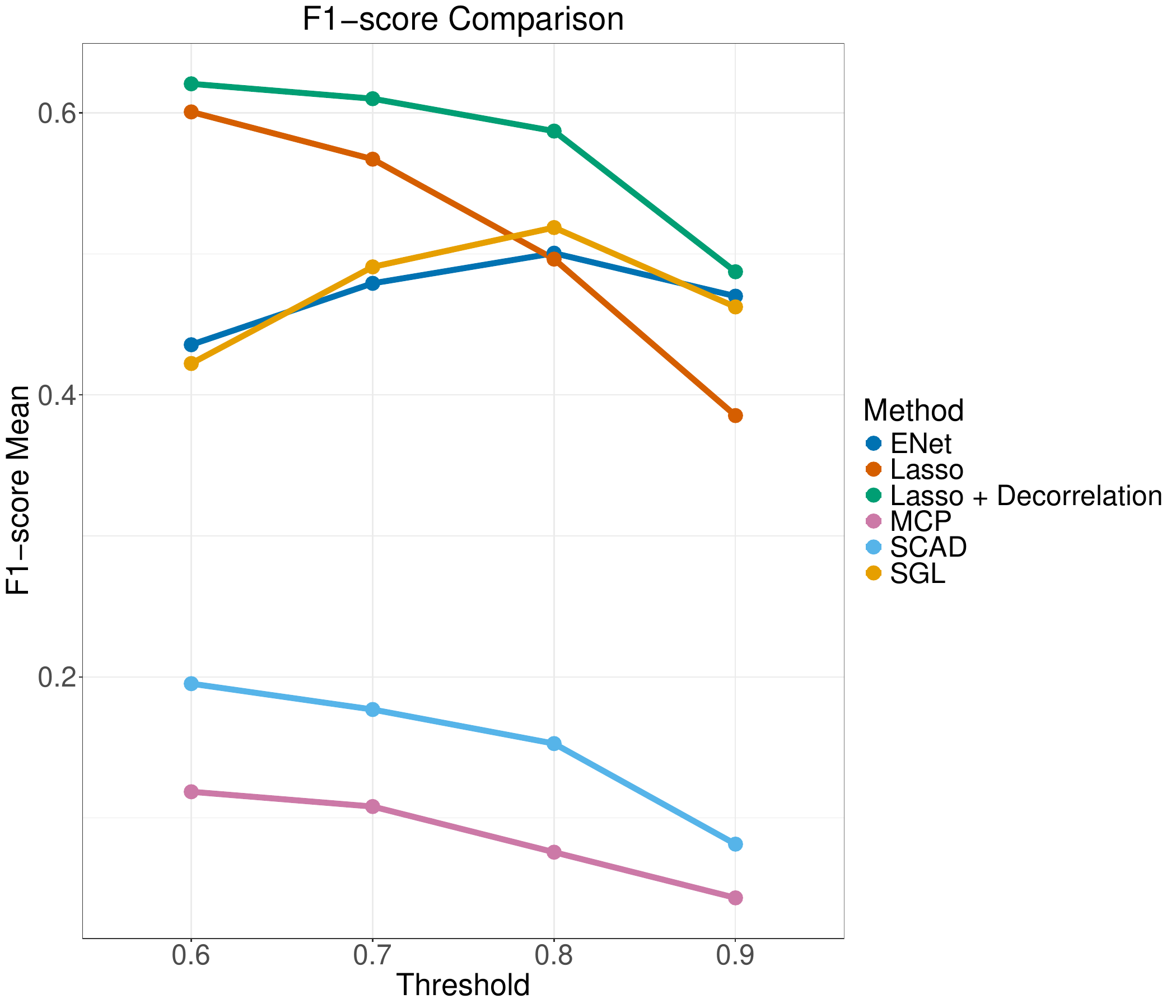}
    \caption{Comparison of F1-score across stability selection with different selection algorithms and decision-making thresholds}
    \label{fig6}
\end{figure}

Figure \ref{fig5} illustrates that Lasso, when combined with our method, outperforms all other candidates in terms of selection stability. Similarly, Figure \ref{fig6} demonstrates that our method achieves superior performance across all decision thresholds with respect to the F1-score. These findings suggest the potential of our method to enhance selection performance not only for the Lasso but possibly for other variable selection methods as well. Although a full exploration of this possibility lies beyond the scope of this paper, supporting numerical results are provided in the appendix.

\subsection*{Riboflavin Data}
As described in Section \ref{s3}, the Riboflavin dataset consists of $p = 4,088$ genes, with the primary objective being to identify the key genes that are associated with the riboflavin production. We set the number of sub-samples to $B = 200$. In contrast to \citet{nouraie2024selection}, the application of our method resulted in the existence of $\lambda_{\text{stable}}$. Figure \ref{fig7:fig1}, introduced in \citet{nouraie2024selection}, indicates that $\lambda_{\text{stable}}$ exists and its corresponding stability values have converged to $0.75$ after almost $50$ sub-samples. The blue shade around the stability line presents the $95\%$ confidence interval for $\hat{\Phi}(M(\lambda_{\text{stable}}))$ values. Under the regularisation value $\lambda_{\text{stable}} = 0.351$ rounded to three decimal places, the genes \texttt{YXLD\_at} and \texttt{LYSC\_at}, with selection frequencies of 0.945 and 0.805 respectively, were identified as relevant genes.

\subsection*{Affymetrix Rat Genome 230 2.0 Array}

\begin{figure}[H]
    \centering
    \subfloat[Riboflavin data\label{fig7:fig1}]{
        \includegraphics[width=0.45\textwidth, height = 0.4\textwidth]{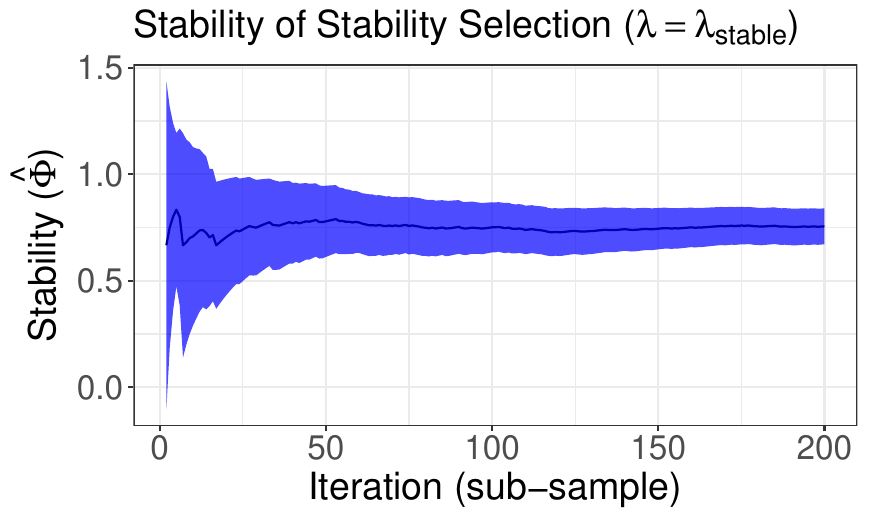}
    }
    \hfill
    \subfloat[Microarray data\label{fig7:fig2}]{
        \includegraphics[width=0.45\textwidth, height = 0.4\textwidth]{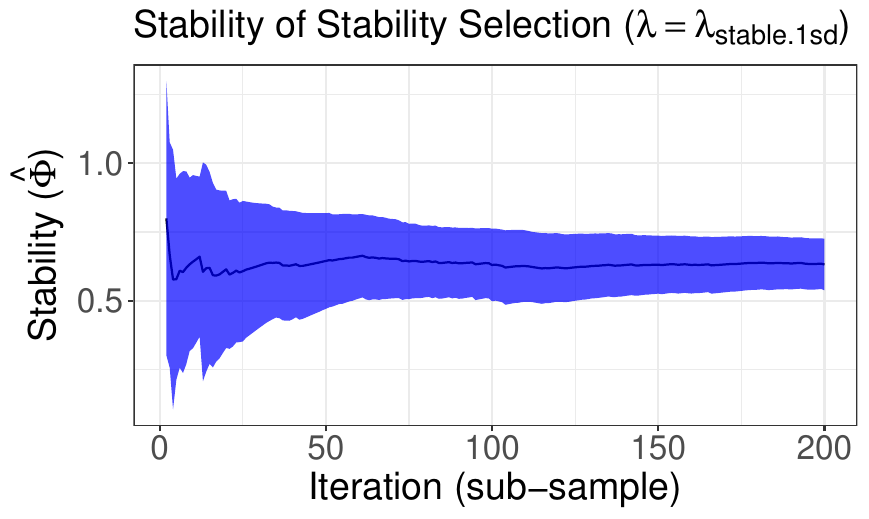}
    }
    \caption{Selection stability of  stability selection with the Lasso over sequential sub-sampling on the real data}
    \label{fig7}
\end{figure}

As mentioned in Section \ref{s3}, the rat microarray data consists of $p = 3,083$ gene probes. The main aim for this data is to identify probes that are associated with the TRIM32 probe. Again, we choose the number of sub-samples $B = 200$. Due to the instability of the outcomes, $\lambda_{\text{stable}}$ does not exist in this problem but, the stability level is much more than what was reported in \citet{nouraie2024selection}. Figure \ref{fig7:fig2} shows that the stability values for $\lambda_{\text{stable-1sd}}$ are converged to around $0.63$. Under the regularisation value $\lambda_{\text{stable-1sd}} = 0.037$ rounded to three decimal places, the probes \texttt{X1389457\_at} and \texttt{X1376747\_at} with selection frequencies of 0.945 and 0.875 respectively, were identified as relevant probes.

Although \citet{nouraie2024selection} recommended $B^{*} = 200$ as a rule of thumb for the stability convergence when applying stability selection with the Lasso, Figure \ref{fig7:fig1} and \ref{fig7:fig2} demonstrate that, with our method, the stability values may converge more rapidly.

\section{Discussion and Conclusion}\label{s5}
Although the classical Gram-Schmidt algorithm is used in this paper, the modified version \citep{golub2013matrix} has been developed to provide better numerical stability. A detailed comparison of these two variants is beyond the scope of this paper.

\citet{meinshausen2010stability} introduced stability selection not only for variable selection, but also for structure estimation tasks such as graphical modelling. In this paper, we demonstrated that decorrelating variables prior to selection enhances stability; however, when the goal is to estimate dependencies among variables—as in graphical models—such decorrelation is not recommended.

\subsection*{Why PCA Is Not Suitable in This Context?}
We now discuss our preference for Gram–Schmidt process over Principal Component Analysis \citep[PCA;][]{PCA}. Although both methods produce orthogonal vectors, their underlying principles differ substantially. PCA operates on the covariance matrix of the data to identify directions that capture maximal variance, with each principal component formed as a linear combination of all original variables. In contrast, the Gram–Schmidt process orthogonalises a set of vectors by iteratively removing projections onto previously obtained vectors, thereby ensuring that each resulting vector remains directly associated with a specific original vector (variable). This property enables the use of the resulting orthonormal matrix $Q$ as a proxy for performing variable selection on the original design matrix $X$. Therefore, although PCA is an effective tool for dimension reduction, it is not appropriate for the purpose of variable decorrelation in our context.

\subsection*{Comparison with Forward Selection}
The Gram–Schmidt process and forward stepwise selection are both iterative procedures, yet they are grounded in fundamentally different principles. The Gram–Schmidt process orthogonalises the variables, producing an orthonormal basis that removes correlation among variables. On the other hand, forward stepwise selection incrementally adds variables based on their correlation with the residuals, with the objective of maximising predictive performance. Notably, Gram–Schmidt identifies informative directions without reference to the response variable, whereas forward selection explicitly targets variables that enhance model fit. However, since we propose to order the variables according to their predictive power, the procedure bears some resemblance to forward selection. Nonetheless, it differs fundamentally in that it does not involve iterative model fitting on the response variable.

\subsection*{Connection to Stable Learning}
Stable learning has emerged as a rapidly growing research area, focused on identifying patterns that remain robust despite variations in the distribution of training and testing data (see, e.g., \citet{Shen_Cui_Zhang_Kunag_2020}, \citet{Zhang_2021_CVPR}, and \citet{cui2022stable}). Given the challenges posed by correlated variables within this field \citep{Shen_Cui_Zhang_Kunag_2020}, we contend that the methodology presented in this paper is closely related to this developing area of research, providing a promising direction for future investigation.

Furthermore, although in this paper the decorrelation procedure is applied to the entire dataset, it may be more appropriate to perform the variable decorrelation separately within each environment, particularly in settings where the data comprise samples from distinct environments, which is a common scenario in the context of stable learning \citep{Shen_Cui_Zhang_Kunag_2020}.

\subsection*{Connection to the Rashomon Sets}
The index-sensitivity of the Gram–Schmidt process, which constructs an orthonormal set of variables based on their ordering, is reminiscent of Rashomon sets—collections of models that yield similarly high predictive performance using different subsets of variables \citep{Rashomon}. Although this topic lies beyond the scope of this paper, it would be of interest to investigate how different orderings of the design matrix lead to distinct predictive models, and how this phenomenon relates to the concept of Rashomon sets, for example in the context of \citet{kissel2024forward}.

\subsection*{Future Work}
Although in this paper the methodology is presented in the context of linear regression, it is evident that correlated variables also pose challenges in non-linear settings \citep{gregorutti2017correlation}. Although addressing this issue is beyond the scope of this paper, we recommend further exploration in this area as a promising direction for future research. Furthermore, although the methodology has been presented within a regularisation framework, we hypothesise that the proposed approach could also be effective for non-regularisation-based methods, such as filter or wrapper methods \citep{guyon2003introduction}, as correlated variables may introduce noise in these methods as well. This suggests an additional promising direction for future research.

\section*{Competing interests}
The authors declare that they have no conflict of interest.

\section*{Author contributions statement}
 
Mahdi Nouraie was responsible for drafting the manuscript, the development of the research methodology and for writing the computer code used throughout. Samuel Muller and Connor Smith provided critical feedback on the content of the manuscript, refining the clarity and scope of the manuscript and the computer code.  

\section*{Data Availability}
The Riboflavin dataset is accessible via the \texttt{hdi} package in R \citep{hdi-package}. The rat microarray data can be obtained from the National Center for Biotechnology Information (NCBI) website at \url{www.ncbi.nlm.nih.gov}, under accession number GSE5680.

The source code used for the paper is accessible through the following GitHub repository: \url{https://github.com/MahdiNouraie/Variable-Decorrelation}. Furthermore, the \texttt{DVS} R package, which facilitates the use of the methodology introduced in this paper, is available through \url{https://github.com/MahdiNouraie/DVS}.

\section*{Acknowledgments}
Mahdi Nouraie was supported by the Macquarie University Research Excellence Scholarship (20213605). Samuel Muller was supported by the Australian Research Council Discovery Project Grant (DP230101908).

\bibliographystyle{plainnat}
\bibliography{citation}

\section*{Appendix}

\subsection*{Gram-Schmidt Process}
\begin{algorithm}[H]
\caption{Gram-Schmidt Orthonormalisation}
\begin{algorithmic}[1]
    \State \textbf{Input:} A matrix \( X \in \operatorname{mat}(n,p) \) with columns \( X_1, X_2, \dots X_p \)
    \State \textbf{Output:} An orthonormal matrix \( Q \in \operatorname{mat}(n,p) \) and an upper triangular matrix \( R \in \operatorname{mat}(p,p) \)
    
    \State \textbf{Initialise:} Set \( Q \) and \( R \) as zero matrices.
    
    \For{$j = 1$ \textbf{to} $p$}
        \State Set \( v = X_j \)
        \If{$j > 1$}
            \For{$i = 1$ \textbf{to} $j-1$}
                \State Compute \( R_{ij} = Q_i^\top X_j \)
                \State Update \( v = v - R_{ij} Q_i \)
            \EndFor
        \EndIf
        \State Compute \( R_{jj} = \| v \|_2 \)
        \State Normalise \( Q_j = \frac{v}{R_{jj}} \)
    \EndFor
\end{algorithmic}\label{alg-Gram}
\end{algorithm}

\subsection*{Proof of Corollary \ref{Corollary-consistency}}
\begin{proof} Here, we denote the ordered design matrix by $X$. Since the Gram-Schmidt process decomposes $X$ as $X = QR$, by substituting $Q$ as the new design matrix, the linear model $\boldsymbol{Y} = X\boldsymbol{\beta} + \boldsymbol{\varepsilon}$ can be rewritten as $\boldsymbol{Y} = Q(R\boldsymbol{\beta}) + \boldsymbol{\varepsilon}$. The irrepresentable condition for this form is $\|Q_{N}^\top Q_{S} (Q_{S}^\top Q_{S})^{-1}\|_{\infty} < 1$. By the orthogonality of the columns of $Q$ and the disjointness of the signal and noise variable sets, we have $Q_{N}^\top Q_{S} = 0 \in \operatorname{mat}(|N| \times |S|)$, and $Q_{S}^\top Q_{S} = \mathbb{I} \in \operatorname{mat}(|S| \times |S|)$, where $|S|$ denotes the cardinality of the signal set $S$. Therefore, it follows that $\|Q_{N}^\top Q_{S} (Q_{S}^\top Q_{S})^{-1}\|_{\infty} = 0 < 1$. \end{proof}

\subsection*{Proof of Corollary \ref{Corollary-stability}}

\begin{proof}
By applying Algorithm \ref{alg-Gram} to $X$ and keeping the matrix $Q$, we know that $Q_{1}$ contributes more to the span of $Q$ than $Q_{2}$. This is because $X_{2}$ must provide some information that is not expressed by $X_{1}$. Since we assumed the two variables are equally relevant to the response variable, this implies that, from our linear modelling perspective, they are essentially the same. Therefore, $Q_{2}$ is merely an orthogonal vector to $Q_{1}$ and does not contribute to the prediction of the response variable $\boldsymbol{Y}$.

Given the orthogonality of $Q$, the Lasso solution treats each variable independently, with each coefficient determined solely by the inner product $Q_j^\top Y$. For $Q_1$ and $Q_3$, these inner products equal $\beta_1$ and $1$, up to noise. Since both values exceed a reasonable regularisation value $\lambda$, the corresponding coefficients are retained in the solution. In contrast, $Q_2$ is not associated with any signal, so $Q_2^\top Y$ contains only noise and remains below the threshold for a reasonable regularisation value $\lambda$. As a result, the model consistently selects only $Q_1$ and $Q_3$.

The variance estimator $s_j^2$ is maximised when $\overline{M(\lambda)_j} = 0.5$ and decreases as $\overline{M(\lambda)_j}$ deviates from 0.5. The selection frequencies of the first two variables using $X$ will converge to 0.5, as established in Corollary 2 of \citet{faletto2022cluster}. In contrast, as demonstrated in the last paragraph, replacing $X$ with $Q$ results in a shift away from this maximum variance for the first two variables. Consequently, $s_j^2$ becomes smaller for $j = 1,2$ when using $Q$. Given that the stability measure defined in Equation~\eqref{eqn: phi} is strictly decreasing in the total selection variance, the corollary is proved.

\end{proof}

\subsection*{Beyond Lasso}

Here, we evaluate our method on some other regularisation-based variable selection models. We use the second simulation scenario where in each group of correlated variables, the last index corresponds to an active variable. For all selection techniques, the regularisation value is determined using $\lambda_{\text{stable-1sd}}$.

\subsubsection*{Elastic Net} 
\begin{figure}[H]
    \centering
    \subfloat[Stability\label{fig8:fig1}]{
        \includegraphics[width=0.45\textwidth, height = 0.4\textwidth]{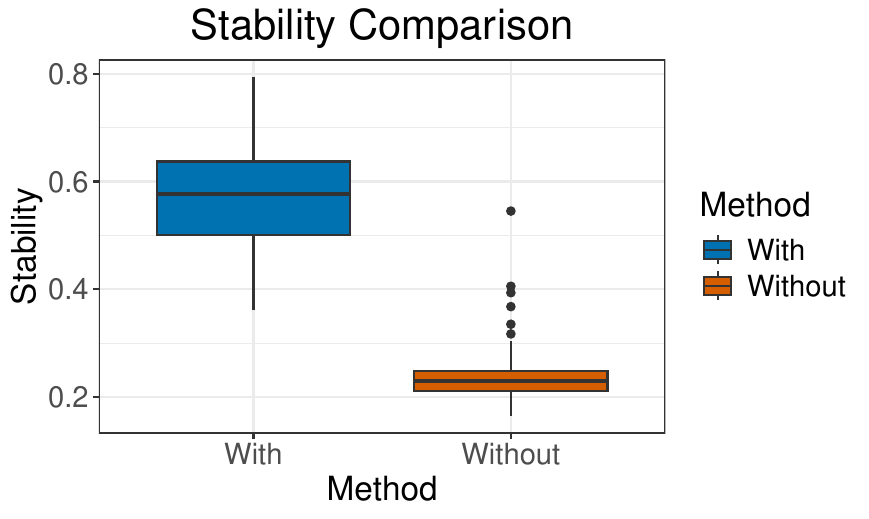}
    }
    \hfill
    \subfloat[F1-score\label{fig8:fig2}]{
        \includegraphics[width=0.45\textwidth, height = 0.4\textwidth]{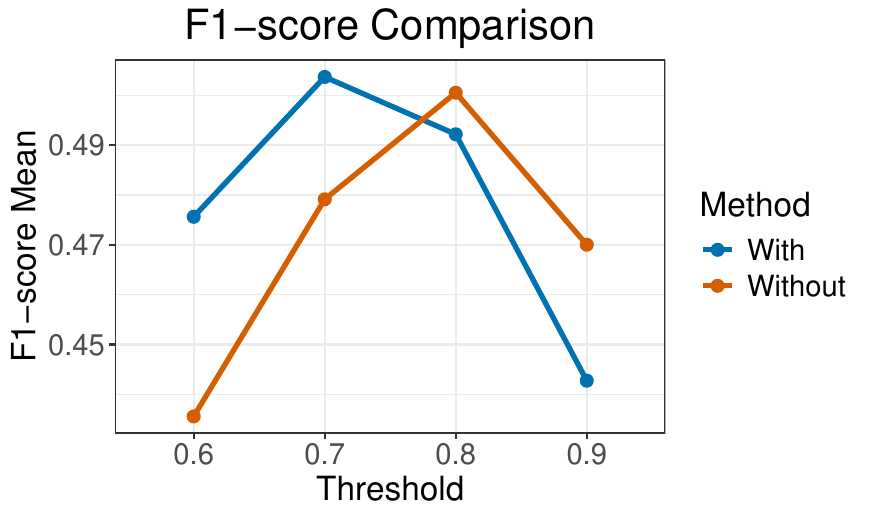}
    }
    \caption{Comparing Stability and F1-score between stability selection using the ENet with and without the variable decorrelation for the second simulation scenario}
    \label{fig8}
\end{figure}

We compare stability selection using the ENet with and without the variable decorrelation. The mixing parameter $\texttt{alpha} = 0.2$. Figure \ref{fig8:fig1} highlights that employing our method significantly enhances the stability of selection results. In addition, Figure \ref{fig8:fig2} illustrates that the F1-score obtained with variable decorrelation surpasses that of the alternative method at lower threshold levels, but becomes inferior at higher thresholds.

\subsubsection*{Smoothly Clipped Absolute
Deviation}

\begin{figure}[H]
    \centering
    \subfloat[Stability\label{fig9:fig1}]{
        \includegraphics[width=0.45\textwidth, height = 0.4\textwidth]{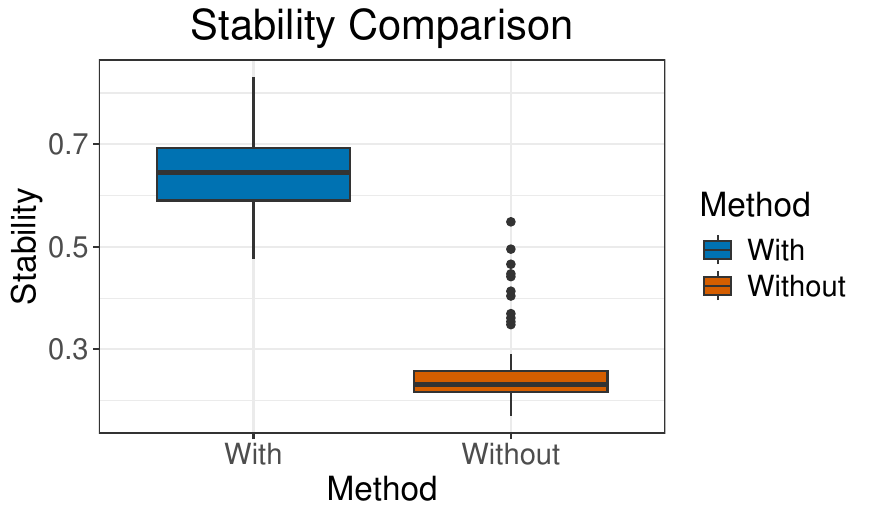}
    }
    \hfill
    \subfloat[F1-score\label{fig9:fig2}]{
        \includegraphics[width=0.45\textwidth, height = 0.4\textwidth]{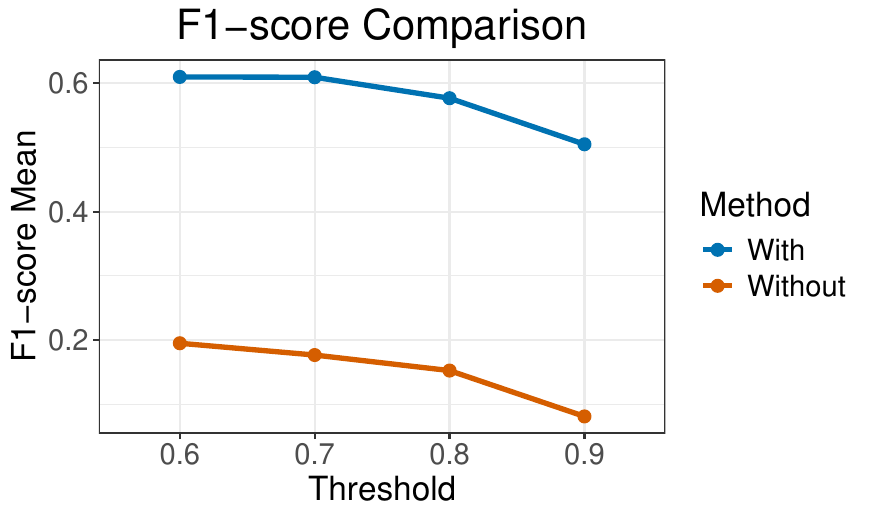}
    }
    \caption{Comparing Stability and F1-score between stability selection using the SCAD with and without the variable decorrelation for the second simulation scenario}
    \label{fig9}
\end{figure}

Figure \ref{fig9} demonstrates that variable decorrelation leads to improvements in both stability and accuracy.

\subsubsection*{Minimax Concave Penalty}

\begin{figure}[H]
    \centering
    \subfloat[Stability\label{fig10:fig1}]{
        \includegraphics[width=0.45\textwidth, height = 0.4\textwidth]{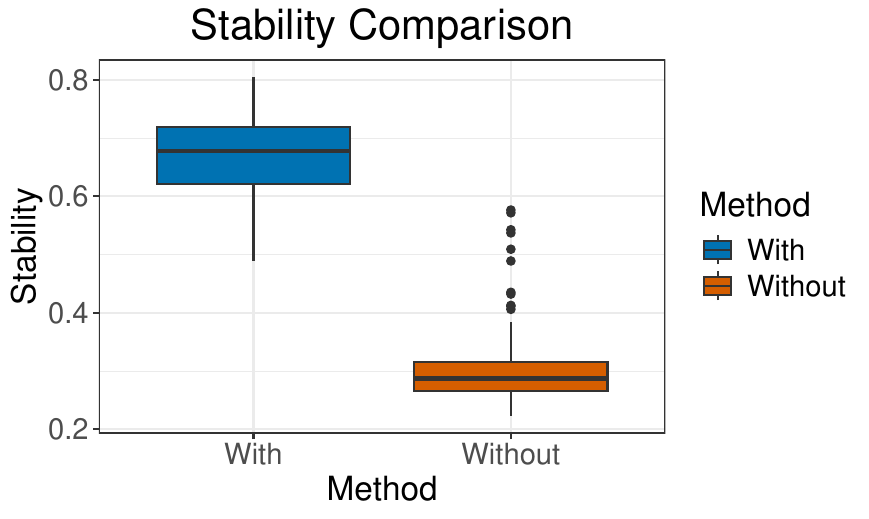}
    }
    \hfill
    \subfloat[F1-score\label{fig10:fig2}]{
        \includegraphics[width=0.45\textwidth, height = 0.4\textwidth]{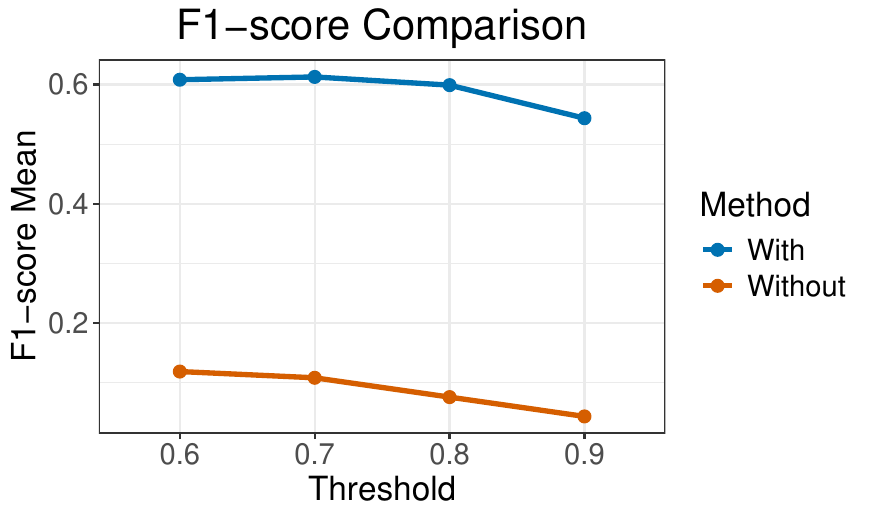}
    }
    \caption{Comparing Stability and F1-score between stability selection using the MCP with and without the variable decorrelation for the second simulation scenario}
    \label{fig10}
\end{figure}

Figure \ref{fig10} presents patterns similar to Figure \ref{fig9}.

\subsubsection*{Sparse Group Lasso}
\begin{figure}[H]
    \centering
    \subfloat[Stability\label{fig11:fig1}]{
        \includegraphics[width=0.45\textwidth, height = 0.4\textwidth]{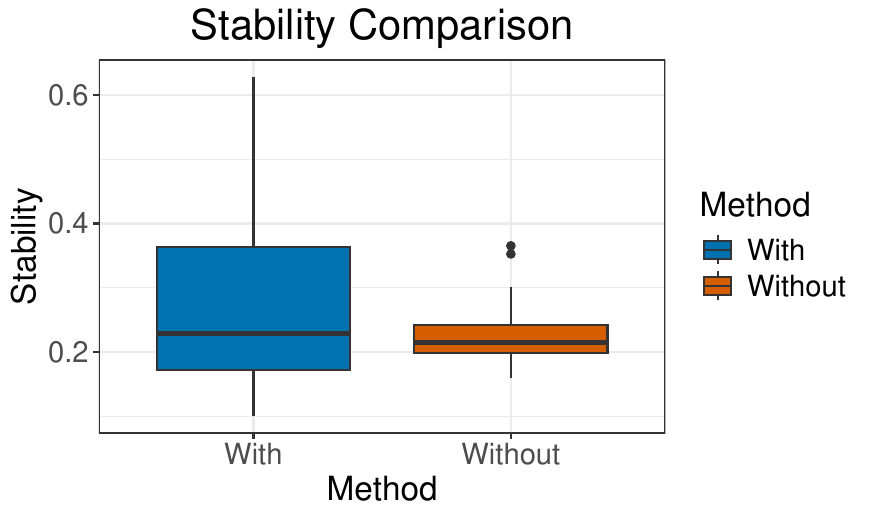}
    }
    \hfill
    \subfloat[F1-score\label{fig11:fig2}]{
        \includegraphics[width=0.45\textwidth, height = 0.4\textwidth]{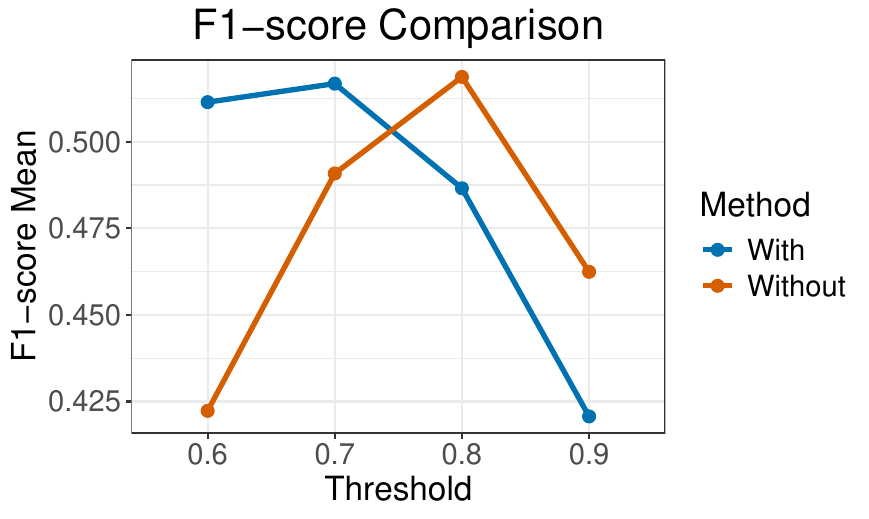}
    }
    \caption{Comparing Stability and F1-score between stability selection using the SGL with and without the variable decorrelation for the second simulation scenario}
    \label{fig11}
\end{figure}

The SGL encourages sparsity both between and within groups of correlated variables. Since the SGL requires the grouping structure of the variables, we provide the grouping index information to the model beforehand. Figure \ref{fig11:fig1} illustrates that our method yields a slight positive effect on the stability of the results when compared to the previous selection techniques. This outcome is expected, as the loss function of the SGL is specifically designed to guide the model through group-related ambiguities by incorporating group indices and addressing sparsity at both the between- and within-group levels. However, it is noteworthy that our method still provides some advantages, even when applied to such a complex model. Figure \ref{fig11:fig2} indicates that the selection F1-score improves when the variable decorrelation is applied and decision-making threshold is low. However, at higher thresholds, the decorrelation negatively affects the F1-score, similar to the pattern observed in Figure \ref{fig8:fig2}.

\end{document}